\documentclass{JHEP3}
\usepackage{epsfig}
\usepackage{amsbsy}
\usepackage{varioref}
\usepackage{pifont}
\usepackage{axodraw}

\newcommand{\hs}{\hspace*{0.5cm}}
\newcommand{\vs}{\vspace*{0.5cm}}
\newcommand{\be}{\begin{equation}}
\newcommand{\ee}{\end{equation}}
\newcommand{\bea}{\begin{eqnarray}}
\newcommand{\eea}{\end{eqnarray}}
\newcommand{\bary}{\begin{array}}
\newcommand{\eary}{\end{array}}
\newcommand{\bit}{\begin{itemize}}
\newcommand{\eit}{\end{itemize}}
\newcommand{\ben}{\begin{enumerate}}
\newcommand{\een}{\end{enumerate}}
\newcommand{\crn}{\nonumber \\}
\newcommand{\nn}{\nonumber}

\newcommand{\al}{\alpha}

\newcommand{\fr}{\frac}

\newcommand{\bc}{\begin{center}}
\newcommand{\ec}{\end{center}}

\def\sla#1{\ifmmode%
\setbox0=\hbox{$#1$}%
\setbox1=\hbox to\wd0{\hss$/$\hss}\else%
\setbox0=\hbox{#1}%
\setbox1=\hbox to\wd0{\hss/\hss}\fi%
#1\hskip-\wd0\box1 }
\title{ Neutralinos and charginos  in supersymmetric
economical 3-3-1 model}
\author{ D. T. Huong and H. N. Long\\
Institute of  Physics, VAST, P. O. Box 429, Bo Ho, Hanoi
10000, Vietnam\\
 E-mail: \email{dthuong@iop.vast.ac.vn},
\email{hnlong@iop.vast.ac.vn}, }

\abstract{Fermion superpartners - neutralinos and charginos in
the supersymmetric economical 3-3-1 model are studied. By
imposition $R$ parity, their masses and eigenstates  are derived.
 Assuming that Bino-like is dark matter, its mass density is calculated.
The cosmological dark matter density gives a bound on  mass  of
LSP neutralino  in the range of 20 $\div$ 100 GeV,  while the
bound on mass of the lightest slepton is  in the range of 60
$\div$ 130 GeV }

\keywords{Supersymmetric models, Supersymmetric partners of known
particles, Models beyond the standard model,  Dark matter }

\begin{document}


\maketitle

\section{\label{intro}Introduction}

The Standard Model (SM) of high energy physics provides a
remarkable successful  description of presently known phenomena.
In spite of these successes, it fails to explain several
fundamental issues like generation number puzzle, neutrino masses
and oscillations, the origin of charge quantization, CP violation,
etc.

One of the simplest solutions to these problems is to enhance the
SM symmetry  $\mathrm{SU}(3)_{C} \otimes \mathrm{SU}(2)_{L}
\otimes \mathrm{U}(1)_{Y}$ to $\mathrm{SU}(3)_{C} \otimes
\mathrm{SU}(3)_{L} \otimes \mathrm{U}(1)_{X}$ (called 3-3-1 for
short)~\cite{ppf,flt,331rh}  gauge group.
 One of the main motivations to study this kind of models
is an explanation in part of the generation number puzzle. In the
3-3-1 models, each generation is not anomaly free; and the model
becomes anomaly free if one of quark families behaves differently
from other two. Consequently, the number of generations is
multiple of the color number. Combining with the QCD asymptotic
freedom, the generation number has to be three. For the neutrino
masses and oscillations, the electric charge quantization and CP
violation issues in the 3-3-1 models, the interested readers can
find in Refs. \cite{neu331}, \cite{chargequan} and \cite{CP331},
respectively.

In one of the 3-3-1 models, the right-handed neutrinos are in
bottom of the lepton triplets \cite{331rh} and three Higgs
triplets are required. It is worth noting that, there are two
Higgs triplets with {\it neutral components in the top and
bottom}. In the earlier version, these triplets can have vacuum
expectation value (VEV) either on the top or in the bottom, but
not in both. Assuming that all neutral components in the triplet
can have VEVs, we are able to reduce number of triplets in the
model to be two ~\cite{ponce,haihiggs} (for a review, see
~\cite{ahep}). Such a scalar sector is minimal, therefore it has
been called the economical 3-3-1 model~\cite{higgseconom}. In a
series of papers, we have developed and proved that this
non-supersymmetric version is consistent, realistic and very rich
in physics \cite{haihiggs,higgseconom, dlhh,dls1}.

In the other hands, due to the ``no-go" theorem of Coleman-Mandula
\cite{nogo}, the internal $G$ and external $P$ spacetime
symmetries can only be {\it trivially} unified. In addition, the
mere fact that the ratio $M_P/M_W$ is so huge is already a
powerful clue to the character of physics beyond the SM, because
of the infamous hierarchy problem. In the framework of new
symmetry called a supersymmetry \cite{susy,martin}, the above
mentioned problems can be solved. One of the intriguing features
of supersymmetric theories is that the Higgs spectrum
(unfortunately, the only part of the SM is still not discovered)
is quite constrained.

It is known that the economical (non-supersymmetric) 3-3-1 model
does not furnish any candidate for self-interaction dark
matter~\cite{SIDM} with the condition given by Spergel and
Steinhardt~\cite{ss}. With a larger content of the scalar sector,
the supersymmetric version is expected to have a candidate for the
self-interaction dark matter.  An supersymmetric version of the
minimal version (without extra lepton) has been constructed in
Ref.~\cite{msusy} and its scalar sector was  studied in
Ref.~\cite{duongma}. Lepton masses in framework of the above
mentioned model was presented in Ref.~\cite{leptonmassm331}, while
potential discovery of supersymmetric particles  was studied in
\cite{consm331}. In Ref.~\cite{longpal}, the $R$ - parity
violating interaction was applied for  instability of the proton.

The supersymmetric version of the 3-3-1 model with right-handed
neutrinos~\cite{331rh} has already been constructed in
Ref.~\cite{s331r}.  The scalar sector was considered in
Ref.~\cite{scalarrhn} and neutrino mass was studied in
Ref.~\cite{marcos}.  Note that there is  three-family  versions in
which lepton families are  treated differently ~\cite{nonst331}
and their supersymmetric versions are presented in
Ref.~\cite{consnst}. It is worth mentioning that in the previous
papers  on supersymmetric version of the 3-3-1  models, the main
attention was given to the gauge boson, lepton mass and  Higgs
sectors.
 An supersymmetric version of the economical 3-3-1 model has been
constructed in Ref. \cite{susyeco}. Some interesting features such
as Higgs bosons with masses equal to that of the gauge bosons --
the $W$ and the bileptons $X$ and $Y$, have been pointed out in
Ref. \cite{higph}. Sfermions have been considered in Ref.
\cite{jhep}.

In a supersymmetric extension of the (beyond) SM, each of the
known fundamental particles must be in either a chiral or gauge
supermultiplet and have a superpartner with spin differing by 1/2
unit. Both gauge and scalar bosons have spin-$\fr 1 2$
superpartners with the electric charges equal to that of their
originals: called neutralinos without  electric charge and
charginos if carrying the latter one. In the Minimal
Supersymmetric Standard Model (MSSM), in some scenario, the
neutralino can be the lightest and plays a role of dark matter. In
this paper, we will focus an attention to  neutralinos and
charginos in the supersymmetric economical 3-3-1 model.

This article is organized as follows. In Sec.~\ref{model} we
present fermion and scalar content in the supersymmetric
economical 3-3-1 model. The necessary parts of Lagrangian is also
given.  In Section \ref{neutral}, we deal with neutralinos sector.
To find eigenstates and their masses, we have to adopt some
assumptions. Section \ref{chargino} is devoted for charginos. In
 Section \ref{neudarkmatt} we present analysis  of
relic neutralino dark matter mass density and the limit on its
mass. Finally, we summarize our results and make conclusions in
the last section - Sec.~\ref{concl}.

\section{A review of the model}
\label{model} In this section we first recapitulate the basic
elements of the supersymmetric economical 3-3-1 model
\cite{susyeco}. $R-parity$ and some constraints on the couplings
are also presented.

\subsection{\label{parcontent}Particle content }

The superfield content in this paper is defined in a standard way
as follows \be \widehat{F}= (\widetilde{F}, F),\hs \widehat{S} =
(S, \widetilde{S}),\hs \widehat{V}= (\lambda,V), \ee where the
components $F$, $S$ and $V$ stand for the fermion, scalar and
vector fields while their superpartners are denoted as
$\widetilde{F}$, $\widetilde{S}$ and $\lambda$, respectively
\cite{susy,s331r}.

The superfield content in the considering model with an
anomaly-free fermionic content transforms under the 3-3-1 gauge
group as
\[
\widehat{L}_{a L}=\left(\widehat{\nu}_{a}, \widehat{l}_{a},
\widehat{\nu}^c_{a}\right)^T_{L} \sim (1,3,-1/3),\hs
  \widehat {l}^{c}_{a L} \sim (1,1,1),
\]
\[
 \widehat Q_{1L}= \left(\widehat { u}_1,\
                        \widehat {d}_1,\
                        \widehat {u}^\prime
 \right)^T_L \sim (3,3,1/3),
 \]
\[
\widehat {u}^{c}_{1L},\ \widehat { u}^{ \prime c}_{L} \sim
(3^*,1,-2/3),\hs \widehat {d}^{c}_{1L} \sim (3^*,1,1/3 ),
\]
\[
\begin{array}{ccc}
 \widehat{Q}_{\alpha L} = \left(\widehat{d}_{\alpha},
 - \widehat{u}_{\alpha},
 \widehat{d^\prime}_{\alpha}\right)^T_{L}
 \sim (3,3^*,0), \hs \al=2,3,
\end{array}
\]
\[
\widehat{u}^{c}_{\alpha L} \sim \left(3^*,1,-2/3 \right),\hs
\widehat{d}^{c}_{\alpha L},\ \widehat{d}^{\prime c}_{\alpha L}
\sim \left(3^*,1,1/3 \right),
\]
where the values in the parentheses denote quantum numbers based
on $\left(\mbox{SU}(3)_C\right.$, $\mbox{SU}(3)_L$,
$\left.\mbox{U}(1)_X\right)$ symmetry.
$\widehat{\nu}^c_L=(\widehat{\nu}_R)^c$ and $a=1,2,3$ is a
generation index. The primes superscript on usual quark types
($u'$ with the electric charge $q_{u'}=2/3$ and $d'$ with
$q_{d'}=-1/3$) indicate that those quarks are exotic ones.

The two superfields $\widehat{\chi}$ and $\widehat {\rho} $ are at
least introduced to span the scalar sector of the economical 3-3-1
model \cite{higgseconom}: \bea \widehat{\chi}&=& \left (
\widehat{\chi}^0_1, \widehat{\chi}^-, \widehat{\chi}^0_2
\right)^T\sim (1,3,-1/3), \crn \widehat{\rho}&=& \left
(\widehat{\rho}^+_1, \widehat{\rho}^0, \widehat{\rho}^+_2\right)^T
\sim  (1,3,2/3). \nn \eea To cancel the chiral anomalies of
higgsino sector, the two extra superfields $\widehat{\chi}^\prime$
and $\widehat {\rho}^\prime $ must be added as follows \bea
\widehat{\chi}^\prime&=& \left (\widehat{\chi}^{\prime 0}_1,
\widehat{\chi}^{\prime +},\widehat{\chi}^{\prime 0}_2
\right)^T\sim ( 1,3^*,1/3), \crn \widehat{\rho}^\prime &=& \left
(\widehat{\rho}^{\prime -}_1,
  \widehat{\rho}^{\prime 0},  \widehat{\rho}^{\prime -}_2
\right)^T\sim (1,3^*,-2/3). \nn \eea

In this model, the $ \mathrm{SU}(3)_L \otimes \mathrm{U}(1)_X$
gauge group is broken via two steps:
 \be \mathrm{SU}(3)_L \otimes
\mathrm{U}(1)_X \stackrel{w,w'}{\longrightarrow}\ \mathrm{SU}(2)_L
\otimes \mathrm{U}(1)_Y\stackrel{v,v',u,u'}{\longrightarrow}
\mathrm{U}(1)_{Q},\label{stages}\ee where the VEVs are defined by
\bea
 \sqrt{2} \langle\chi\rangle^T &=& \left(u, 0, w\right), \hs \sqrt{2}
 \langle\chi^\prime\rangle^T = \left(u^\prime,  0,
 w^\prime\right),\label{ct1}\\
\sqrt{2}  \langle\rho\rangle^T &=& \left( 0, v, 0 \right), \hs
\sqrt{2} \langle\rho^\prime\rangle^T = \left( 0, v^\prime,  0
\right).\nn\eea The VEVs $w$ and $w^\prime$ are responsible for
the first step of the symmetry breaking while $u,\ u^\prime$ and
$v,\ v^\prime$ are for the second one. Therefore, they have to
satisfy the constraints:
 \be
 u,\ u^\prime,\ v,\ v^\prime
\ll w,\ w^\prime. \label{contraint}\ee

 It is emphasized that the VEV structure in (\ref{ct1}) is not
only the key to reduce Higgs sector but also the reason for
complicated  mixing among gauge, Higgs bosons, etc. As it will be
shown in the following, the mentioned VEV structure causes flavour
violation in the $D$-term contributions.

The vector superfields $\widehat{V}_c$, $\widehat{V}$ and
$\widehat{V}^\prime$ containing the usual gauge bosons are,
respectively, associated with the $\mathrm{SU}(3)_C$,
$\mathrm{SU}(3)_L$ and $\mathrm{U}(1)_X $ group factors. The
colour and flavour vector superfields have expansions in the
Gell-Mann matrix bases $T^a=\lambda^a/2$ $(a=1,2,...,8)$ as
follows\bea \widehat{V}_c &=& \fr{1}{2}\lambda^a
\widehat{V}_{ca},\hs
\widehat{\overline{V}}_c=-\fr{1}{2}\lambda^{a*}
\widehat{V}_{ca};\hs \widehat{V} = \fr{1}{2}\lambda^a
\widehat{V}_{a},\hs \widehat{\overline{V}}=-\fr{1}{2}\lambda^{a*}
\widehat{V}_{a},\nn\eea where an overbar $^-$ indicates complex
conjugation. For the vector superfield associated with
$\mathrm{U}(1)_X$, we normalize as follows \[ X \hat{V}'= (XT^9)
\hat{B}, \hs T^9\equiv\fr{1}{\sqrt{6}}\mathrm{diag}(1,1,1).\] The
gluons are denoted by $g^a$ and their respective gluino partners
by $\lambda^a_{c}$, with $a=1, \ldots,8$. In the electroweak
sector, $V^a$ and $B$ stand for the $\mathrm{SU}(3)_{L}$ and
$\mathrm{U}(1)_{X}$ gauge bosons with their gaugino partners
$\lambda^a_{V}$ and $\lambda_{B}$, respectively.

With the superfields as given, the full Lagrangian is defined by
$\mathcal{L}_{susy}+\mathcal{L}_{soft}$, where the first term is
supersymmetric part, whereas the last term breaks explicitly the
supersymmetry \cite{susyeco}. The interested reader can find more
details on this Lagrangian in the above mentioned article. In the
following, only terms relevant to our calculations are displayed.

\subsection{$R$-parity} For the further analysis, it is convenience
to introduce $R$-parity in the model.  Following Ref.
\cite{marcos}, $R$-parity can be expressed as follows
\begin{equation}
R-parity=(-1)^{2S}(-1)^{3({\mathcal B}+{\mathcal L})}
\label{rfor}\end{equation} where invariant charges ${\mathcal L}$
and ${\mathcal B}$ (for details, see  Ref. \cite{changlong}) are
given by~\cite{jhep}
\begin{equation}
\begin{array}{|c|c|c|c|c|}
\hline
  Triplet & L & Q_{1} & \chi  & \rho \\
  \hline
  {\mathcal B} \,\  charge & 0 & \frac{1}{3} & 0  & 0 \\ \hline
  {\mathcal L} \,\  charge & \frac{1}{3} & - \frac{2}{3} & \frac{4}{3}
  & - \frac{2}{3} \\ \hline
\end{array}
\end{equation}
\begin{equation}
\begin{array}{|c|c|c|c|}
\hline
  Anti-Triplet & Q_{\alpha} & \chi^{\prime}  & \rho^{\prime} \\
  \hline
  {\mathcal B} \,\  charge & \frac{1}{3} & 0 &  0 \\ \hline
  {\mathcal L} \,\  charge &  \frac{2}{3} & - \frac{4}{3}
  & \frac{2}{3} \\ \hline
\end{array}
\end{equation}
\begin{equation}
\begin{array}{|c|c|c|c|c|c|}
\hline
  Singlet & l^{c} & u^{c} & d^{c} & u^{\prime c} & d^{\prime c} \\
  \hline
  {\mathcal B} \,\ charge & 0 &- \frac{1}{3} &
  - \frac{1}{3} & - \frac{1}{3} & - \frac{1}{3} \\ \hline
  {\mathcal L} \,\ charge & - 1 & 0 &  0 & 2 & -2 \\ \hline
\end{array}
\end{equation}

\section{The neutralinos sector}
\label{neutral}

 The higginos and electroweak gauginos mix  each with other due to
  effects of the electroweak symmetry breaking. The neutral higginos and
  gauginos combine to make the mass eigenvectors called neutralinos. In this
section,  the mass spectrum and mixing of the neutralinos is
considered.

 \hs The gauginos mass terms come directly from the soft term
  given by
 \be
 \mathcal{L}_{Soft}= \sum_{b=1}^{8}M_b\widetilde{\mathcal{W}}^b
 \widetilde{\mathcal{W}}^b +M_{\widetilde{\mathcal{B}}}\widetilde{\mathcal{B}}
 \widetilde{\mathcal{B}}.
 \label{an1}\ee
Because of the R-parity conservation, the higginos mixing terms
come from the $\mu-$term determined as \be
 \mathcal{L}_{\mu-term}=\mu_{\chi}\widehat{\chi}\widehat{\chi}^\prime
 +\mu_{\rho}\widehat{\rho}\widehat{\rho}^\prime.
 \label{an2}\ee
 Finally, the mixing terms between higginos and gauginos are a result
 of Higgs-higginos-gauginos couplings
 \be
 \mathcal{L}=-\sqrt{2}g\left( \phi^*T^a \psi\right)\lambda^a
 -\sqrt{2}g\lambda^{+a}\left( \psi^+ T^a\phi\right).
 \label{an3}\ee
  Expanding Eqs (\ref{an1}), (\ref{an2}) and (\ref{an3}), we
  obtain the neutralino mass matrix
  in the gauge-eigenatates  basis $\psi^o= \left(\widetilde{\chi^o_1 }
  ,  \widetilde{\chi^{o\prime}_1 }, \widetilde{\chi^o_2 }
  , \widetilde{\chi^{o\prime}_2 }, \widetilde{\rho^o_1},
  \widetilde{\rho^{o\prime}_1}, \widetilde{\mathcal{B}},
  \widetilde{\mathcal{W}_3},
  \widetilde{\mathcal{W}_8}, \widetilde{\mathcal{X}},
  \widetilde{\mathcal{X}^*}\right)$, which is given in
   the Lagrangian form
  \be
  \mathcal{L}=\left(\widetilde{\psi^{o} }\right)^T
  M_{\widetilde{N}}\widetilde{\psi^o}
  \ee
  with the following notations
   \be
  \widetilde{ \mathcal{X}}=\frac{\widetilde{\mathcal{W}}_4
  +i\widetilde{\mathcal{W}}_5}{2},
  \widetilde{ \mathcal{X^*}}=\frac{\widetilde{\mathcal{W}}_4
  -i\widetilde{\mathcal{W}}_5}{2}
   \ee
   and
   \be
M_{\widetilde{N}}=\left(
 \begin{array}{ccccccccccc}
0 & -\mu_\chi & 0 & 0 & 0 & 0 & -\frac{g^\prime u}{3\sqrt{6}}
 & \frac{gu}{2} & \frac{g u}{2\sqrt{3}} & \frac{gw}{\sqrt{2}} & 0 \\
 -\mu_\chi  & 0 & 0& 0& 0 & 0 & \frac{g^\prime u^\prime}{3\sqrt{6}}
 & \frac{gu^\prime}{2}& \frac{g u^\prime}{2\sqrt{3}} & \frac{gw^\prime}
 {\sqrt{2}} & 0 \\
 0 & 0 & 0 &  &-\mu_\chi & 0  &- \frac{g^\prime w}{3\sqrt{6}}
  & 0& -\frac{gw}{\sqrt{3}} & 0&
  \frac{gu}{\sqrt{2}}\\
  0& 0& -\mu_\chi  &0 & 0 & 0 & \frac{g^\prime w^\prime}{3\sqrt{6}}
  & 0 & -\frac{gw^\prime}{\sqrt{3}} & 0 & \frac{gu^\prime}{\sqrt{2}} \\
 0 & 0 & 0 & 0 & 0 & -\mu_\rho & \frac{2g^\prime v}{3\sqrt{6}} &
 -\frac{gv}{2} & \frac{gv}{2\sqrt{3}} & 0 & 0 \\
 0 & 0 & 0 & 0 & -\mu_\rho & 0 & -\frac{2g^\prime v^\prime}{3\sqrt{6}}
 & -\frac{gv^\prime}{2} & \frac{gv^\prime}{2\sqrt{3}} & 0& 0 \\
 -\frac{g^\prime u}{3\sqrt{6}}& \frac{g^\prime u^\prime}{3\sqrt{6}} &
 -\frac{g^\prime w}{3\sqrt{6}}
 &  \frac{g^\prime w^\prime}{3\sqrt{6}} &  \frac{2g^\prime v}{3\sqrt{6}} &
 -\frac{2g^\prime v^\prime}{3\sqrt{6}} & \mathcal{M_B} & 0 & 0 & 0 & 0 \\
  \frac{gu}{2} & \frac{gu^\prime}{2} & 0 & 0& -\frac{gv}{2}
  & -\frac{gv^\prime}{2} & 0& \mathcal{M}_3 & 0 & 0 & 0 \\
\frac{gu}{2\sqrt{3}} & \frac{gu^\prime}{2\sqrt{3}} &
-\frac{gw}{\sqrt{3}} & -\frac{gw^\prime}{\sqrt{3}} &
\frac{gv}{2\sqrt{3}}  &  \frac{gv^\prime}{2\sqrt{3}}
 & 0 & 0& \mathcal{M}_8& 0 & 0 \\
\frac{gw}{2} & \frac{gw^\prime}{2} & 0 & 0 & 0 & 0 & 0 & 0
& 0 & \mathcal{M}_{45} & 0 \\
0 & 0 & \frac{gu}{2} & \frac{gu^\prime}{2} & 0 & 0 & 0& 0 & 0 & 0
& \mathcal{M}_{45}
        \end{array}
                  \right)\crn
                 \label{an4}\ee
 where $\mathcal{M}_{4} = \mathcal{M}_{5} \equiv
 \mathcal{M}_{45}$.
  The mass matrix $M_{\widetilde{N}}$ can be diagonalized by an
  unitary matrix $U$ to get the mass eigenstates. It means that we
  can find matrix $U$ satisfying:
   \bea
   U M U^{-1} & = & \textrm{Diag}(m_{\widetilde{N}_1},m_{\widetilde{N}_2},m_{\widetilde{N}_3}
   m_{\widetilde{N}_4},m_{\widetilde{N}_5},m_{\widetilde{N}_6},m_{\widetilde{N}_7},\crn
   &&
   m_{\widetilde{N}_8},m_{\widetilde{N}_9},m_{\widetilde{N}_{10}},m_{\widetilde{N}_{11}})
   \eea
   with real positive entries on the diagonal.

    In general, the parameters $\mathcal{M}_B,\mathcal{M}_3,
   \mathcal{M}_8,\mathcal{M}_{45},\mu_\chi,\mu_\rho$ can take
   arbitrary complex phase. However we can choose a convention to
   make $\mathcal{M}_B,\mathcal{M}_3,
   \mathcal{M}_8,\mathcal{M}_{45}$ to be  all real and positive. If we
   choose the parameter $\mu_\chi, \mu_\rho$ to be real and positive
   then we must pick up the $\left \langle \chi \right\rangle,
   \left \langle \chi^\prime \right\rangle,\left \langle \rho \right\rangle,
   \left \langle \rho^\prime \right\rangle$ to be real and
   positive too. If $\mu_\chi $ and $\mu_\rho $ are not real, then we
   obtain the CP violating effects in the potential. Therefore, as
   the same  as in the MSSM ~\cite{martin}, it is convinience to choose the $\mu_\chi, \mu_\rho$
   to be  real but without fixing  the sign of $\mu_\chi, \mu_\rho$.

    Getting  exact  eigenvalues and eigenstates of the mixing
   mass matrix (\ref{an4}) is very difficult task. Hence,  we  make some
   assumptions which is suitable for theoretical comments; and their correctness
   could  be tested by the future experiments.

    In this paper, we assume that
   \bea
   v,v^\prime, u, u^\prime, w, w^\prime\ll \left|\mu_{\rho}-\mathcal{M}_\mathcal{B}
   \right|,\left|\mu_{\rho}-\mathcal{M}_3 \right|,\left|\mu_{\rho}-\mathcal{M}_8
   \right|,\left|\mu_{\rho}-\mathcal{M}_{45} \right|
\label{an55}
   \eea
and
   \bea
   v,v^\prime, u, u^\prime, w, w^\prime\ll \left|\mu_{\chi}-\mathcal{M}_\mathcal{B}
   \right|,\left|\mu_{\chi}-\mathcal{M}_3 \right|,\left|\mu_{\chi}-\mathcal{M}_8
   \right|,\left|\mu_{\chi}-\mathcal{M}_{45} \right|. \label{an5}
   \eea
   In the above limit, using a small perturbation on the
   neutralinos mass matrix (\ref{an4}), we can obtain
   the neutralino mass eigenstates, which  are nearly a ``higginos-like", a
   ``Bino-like",    a ``zino-like", an ``extrazino-like ",  a  ``xino-like",
   and the conjugated of  the ``xino-like"  corresponding to
\bea && \widetilde{N}_1=\widetilde{\mathcal{B}},
\widetilde{N}_2=\widetilde{\mathcal{W}}_3,\widetilde{N}_3=\widetilde{\mathcal{W}}_8,
\widetilde{N}_{4}=\widetilde{\mathcal{X}^*},
\widetilde{N}_{5}=\widetilde{\mathcal{X}},\crn &&
\widetilde{N}_6,\widetilde{N}_7=\frac{\widetilde{\rho}^o
\pm\widetilde{\rho^{\prime o}_1}}{\sqrt{2}},
\widetilde{N}_8,\widetilde{N}_9=\frac{\widetilde{\chi^o}_1
\pm\widetilde{\chi^{o\prime}_1}}{\sqrt{2}},\widetilde{N}_{10},\widetilde{N}_{11}
=\frac{\widetilde{\chi^o}_2 \pm\widetilde{\chi^{\prime
o}_2}}{\sqrt{2}}\eea  with the mass eigenvalues:
 \bea m_{\widetilde{N}_1} &=&\mathcal{M}_\mathcal{B}+\frac{g^{\prime
2}\left[\left(u+u^\prime \right)^2+\left(w+w^\prime
\right)^2\right]}{108\left(\mathcal{M}_\mathcal{B}
+\mu_\chi\right)}+\frac{g^{\prime 2}\left[\left(u-u^\prime
\right)^2+\left(w-w^\prime
\right)^2\right]}{108\left(\mathcal{M}_\mathcal{B}
-\mu_\chi\right)} \crn & & + \frac{g^{\prime 2}\left(v-v^\prime
\right)^2}{27\left(\mathcal{M}_\mathcal{B} -\mu_\rho\right)}+
\frac{g^{\prime 2}\left(v+v^\prime
\right)^2}{27\left(\mathcal{M}_\mathcal{B} +\mu_\rho\right)},\crn
m_{\widetilde{N}_2}&=& \mathcal{M}_{3}+\frac{g^2\left(u-u^\prime
\right)^2}{8\left(\mathcal{M}_3+\mu_{\chi}\right)}+\frac{g^2\left(u+u^\prime
\right)^2}{8\left(\mathcal{M}_3-\mu_{\chi}\right)}\crn &&
+\frac{g^2\left(v+v^\prime
\right)^2}{8\left(\mathcal{M}_3-\mu_{\rho}\right)}+\frac{g^2\left(v-v^\prime
\right)^2}{8\left(\mathcal{M}_3+\mu_{\rho}\right)}, \crn
m_{\widetilde{N}_3}&=& \mathcal{M}_{8}+
\frac{g^2\left[\left(u-u^\prime\right)^2+4
\left(w-w^\prime\right)^2\right]}{24\left(
\mathcal{M}_8+\mu_\chi\right)}+\frac{g^2\left[\left(u+u^\prime\right)^2+4
\left(w+w^\prime\right)^2\right]}{24\left(
\mathcal{M}_8-\mu_\chi\right)} \crn & & +
\frac{g^2\left(v-v^\prime\right)^2}{24\left(
\mathcal{M}_8+\mu_\rho\right)}+\frac{g^2\left(v+v^\prime\right)^2}{24\left(
\mathcal{M}_8-\mu_\rho\right)}, \crn
  m_{\widetilde{N}_{4}}&=&
\mathcal{M}_{45} +\frac{g^2\left[ 2\mu_{\chi}uu^\prime
+\mathcal{M}_{45}\left( u^2+u^{\prime
2}\right)\right]}{2\left(\mathcal{M}_{45}^2
-\mu_{\chi}^2\right)},\crn
 m_{\mathcal{\widetilde{N}}_{5}}&=&
\mathcal{M}_{45}+\frac{g^2\left[ 2\mu_{\chi}ww^\prime
+\mathcal{M}_{45}\left( w^2+w^{\prime
2}\right)\right]}{2\left(\mathcal{M}_{45}^2
-\mu_{\chi}^2\right)},\crn m_{\widetilde{N}_6}
&=&\left|\mu_\rho\right|
 +\frac{g^2\left(v-v^\prime\right)^2}{8\left(\mu_\rho-\mathcal{M}_3\right)}
+\frac{g^2\left(v-v^\prime\right)^2}{24\left(\mu_\rho-\mathcal{M}_8\right)}
+\frac{g^{\prime2}\left(v+v^\prime\right)^2}{27\left
(\mu_\rho-\mathcal{M}_\mathcal{B}\right)},\crn
  m_{\widetilde{N}_7} &=&\left|\mu_\rho\right|
+\frac{g^2\left(v+v^\prime\right)^2}{8\left(\mu_\rho-\mathcal{M}_3\right)}
+
\frac{g^2\left(v+v^\prime\right)^2}{24\left(\mu_\rho-\mathcal{M}_8\right)}
+\frac{g^{\prime2}\left(v-v^\prime\right)^2}{27\left
(\mu_\rho-\mathcal{M}_\mathcal{B}\right)},\crn
 m_{\widetilde{N}_8}
&=&\left|\mu_\chi\right| + \frac{1}{2}\left[
m_{a11}+m_{a22}-\sqrt{\left(m_{a11}-m_{a22}\right)^2+4m_{a12}^2}\right],
\crn  m_{\widetilde{N}_9} &=&\left|\mu_\chi\right| +
\frac{1}{2}\left[
m_{b11}+m_{b22}-\sqrt{\left(m_{b11}-m_{b22}\right)^2+4m_{b12}^2}\right],
\crn  m_{\widetilde{N}_{10}} &=&\left|\mu_\chi\right|
+\frac{1}{2}\left[
m_{a11}+m_{a22}+\sqrt{\left(m_{a11}-m_{a22}\right)^2+4m_{a12}^2}\right],
\crn  m_{\widetilde{N}_{11}} &=&\left|\mu_\chi\right| +
\frac{1}{2}\left[
m_{b11}+m_{b22}+\sqrt{\left(m_{b11}-m_{b22}\right)^2+4m_{b12}^2}\right]
 \eea
 where \bea m_{a11} &=& \frac{1}{126}\left[ \frac{-2g^{\prime
2}\left(
u-u^\prime\right)^2}{\mathcal{M}_\mathcal{B}-\mu_\chi}+9g^2
\left(\frac{3}{\mu_\chi-\mathcal{M}_3}+\frac{1}{\mu_\chi
-\mathcal{M}_8}\right)\left( u+u^\prime \right)^2\right] \nonumber
\\ & & -\frac{3g^2\left(w+w^\prime
\right)^2}{7\left(M_{45}-\mu_\chi \right)}, \crn m_{a12}
&=&\frac{-g^{\prime2}\left(u-u^\prime
\right)\left(w-w^\prime\right)}
{108\left(\mathcal{M}_\mathcal{B}-\mu_\chi\right)}
+\frac{g^2\left(u+u^\prime
\right)\left(w+w^\prime\right)}{12\left(\mathcal{M}_8-\mu_\chi\right)},
\crn m_{a22}&=&-\fr{g^2}{12\left(\mathcal{M}_8-
\mu_\chi\right)\left(\mathcal{M}_{45}-\mu_\chi\right)}
\left\{3\mathcal{M}_8\left(u
+u^\prime\right)^2+2\mathcal{M}_{45}\left(
w+w^\prime\right)^2\right.  \crn  && \left. -\mu_\chi\left[
3\left(u+u^\prime
\right)^2+\left(w+w^\prime\right)^2\right]\right\}
 -\frac{1}{108}\frac{g^{\prime
2}\left(w-w^\prime\right)^2} {\mathcal{M}_\mathcal{B}-\mu_{\chi}},
\crn m_{b11}&=&-\frac{1}{108}\frac{g^{\prime
2}\left(u+u^\prime\right)^2} {\mathcal{M}_\mathcal{B}+\mu_{\chi}}-
\frac{g^2\left( w-w^\prime\right)^2}{4\left(
\mathcal{M}_{45}+\mu_{\chi}\right)}\crn &&
-\frac{g^2}{26}\left(\frac{3}{\mathcal{M}_3+\mu_\chi}+\frac{1}{\mathcal{M}_8+\mu_\chi}
\right)\left(u-u^\prime\right)^2, \crn
m_{b12}&=&\frac{g^2\left(u-u^\prime\right)\left
(w-w^\prime\right)}{12\left(\mathcal{M}_8+\mu_\chi\right)}-\frac{g^{\prime
2}\left( u+u^\prime\right)\left(
w+w^\prime\right)}{108\left(\mathcal{M}_\mathcal{B}+\mu_\chi\right)},
\crn
m_{b22}&=&-\frac{g^2}{12\left(\mathcal{M}_8+\mu_\chi\right)\left(\mathcal{M}_{45}
+\mu_{\chi}\right)}\left\{
 \mu_\chi\left[3\left( u-u^\prime\right)^2 +2\left(
w-w^\prime\right)^2\right]\right. \crn && \left.
+3\mathcal{M}_8\left(u-u^\prime \right)^2+2M_{45}\left(w-w^\prime
\right)^2 \right\} - \frac{g^{\prime 2}\left(w+w^\prime
\right)^2}{\mathcal{M}_\mathcal{B}+\mu_\chi}.
 \eea
  We emphasize that $\mathcal{M}_\mathcal{B},\mathcal{M}_3,\mathcal{M}_8,\mathcal{M}_{45}$
  were   taken real and positive and $\mu_\chi, \mu_\rho$ are real with  arbitrary sign.
   The mass values depend on the numerical values of the
  parameters. In particular case, we assume
  $\mathcal{M}_\mathcal{B}<\mathcal{M}_3<\mathcal{M}_8<\mathcal{M}_{45}\ll\mu_\chi,\mu_\rho$.
  In this case, we obtain the neutralino lightest supersymmetric particle
  (LSP), which  is a Bino-like $\widetilde{N}_1$. In the
  following, we  will focus our attention to  the neutralino LSP.

\section{The charginos sector }
\label{chargino}

 The charged winos  $(
{\widetilde{\mathcal{W}}}^{+},\widetilde{\mathcal{W}}^{-},\widetilde{\mathcal{Y}}^{+},
\widetilde{\mathcal{Y}}^{-})$  mix with  the charged higginos
$(\widetilde{\chi}^{-}$, $\widetilde{\chi}^{\prime +}$, $
\widetilde{\rho_1}^{+}$, $\widetilde{\rho_2}^{+}$,
$\widetilde{\rho_1}^{\prime -}$, $\widetilde{\rho_2}^{\prime -} )$
to form the eigenstates with the electric charges $\pm 1$. They
are called charginos. As  the same as in the MSSM, we will denote
the charginos eigenstates by $C_{i}^{\pm}$. The entries of the
elements in the charginos mass matrix come from $\left( \ref{an1}
\right),\left( \ref{an2} \right)$ and $\left( \ref{an3} \right)$.
In the gauge-eigenstate  basis $\psi^{\pm}=(\widetilde{\mathcal{
W}}^+$, $\widetilde{\mathcal{Y}}^+$, $\widetilde{\rho_1}^+$, $
\widetilde{\rho_2}^+$, $\widetilde{\chi}^{\prime +}$,
$\widetilde{\mathcal{ W}}^-$, $\widetilde{\mathcal{Y}}^-$,
$\widetilde{\rho_1}^{\prime -}$, $\widetilde{\rho_2}^{\prime -}$,
$\widetilde{\chi}^{-})$, the chargino mass terms in the Lagrangian
form are given by
 \be \mathcal{L}_{chargino mass} = \left(\widetilde{\psi}^{\pm}\right)^+
 M_{\widetilde{\psi}} \widetilde{\psi}^{\pm} + H.c\ee
 with the $M_{\widetilde{\psi}}$ having the $2\times2$ block form:
 \be
 M_{\widetilde{\psi}} = \left(%
\begin{array}{cc}
  0 & \mathcal{M }\\
  \mathcal{M}^T & 0 \\
\end{array}%
\right),
 \ee
 where $\mathcal{M}$ is $5 \times 5$ matrix  given by
 \be
 \mathcal{M}=\left(%
\begin{array}{ccccc}
  \mathcal{M}_\mathcal{W} & 0 &\frac{gv^\prime}{\sqrt{2}} & 0 & \frac{gu}{\sqrt{2}}\\
  0 &  \mathcal{M}_{\mathcal{Y}} & 0&\frac{gv^\prime}{\sqrt{2}} &\frac{gw}{\sqrt{2}} \\
  \frac{gv}{\sqrt{2}} & 0 & \mu_\rho & 0 & 0\\
  0 & \frac{gv}{\sqrt{2}} & 0 & \mu_{\rho} & 0 \\
  \frac{gu^\prime}{\sqrt{2}} & \frac{gw^\prime}{\sqrt{2}} & 0 & 0 & \mu_\chi\\
\end{array}%
\right).
 \ee
 In principle, the mixing matrix for positive charged left-handed
fermions and negative charged left-handed fermions are different.
Therefore, we can find two unitary $5 \times 5$ matrices U and V
to relate the gauge eigenstates with the mass eigenstates \be
\left( \begin{array}{c}
 \widetilde{ C}_1^+ \\
  \widetilde{ C}_2^+ \\
  \widetilde{ C}_3^+ \\
  \widetilde{ C}_4^+ \\
  \widetilde{ C}_5^+ \\
\end{array}\right)=V\left(%
\begin{array}{c}
  \widetilde{W}^+ \\
  \widetilde{Y}^+ \\
  \rho_1^+ \\
 \rho_2^+ \\
  \chi^{\prime +} \\
\end{array}%
\right), \left( \begin{array}{c}
 \widetilde{ C}_1^-\\
  \widetilde{ C}_2^- \\
  \widetilde{ C}_3^- \\
  \widetilde{ C}_4^- \\
  \widetilde{ C}_5^- \\
\end{array}\right)=U\left(%
\begin{array}{c}
  \widetilde{W}^- \\
  \widetilde{Y}^- \\
  \rho_1^{\prime -} \\
 \rho_2^{\prime -} \\
  \chi^{-} \\
\end{array}%
\right). \ee This  means that the charginos mass matrix can be
diagonalized
 by two unitary matrices U and V to obtain mass eigenvalues
 \be
 U^*\mathcal{M }V^{-1}=\left(%
\begin{array}{ccccc}
  m_{\widetilde{C}_1} & 0 & 0&0 & 0 \\
  0 &  m_{\widetilde{C}_2} & 0 & 0 & 0 \\
  0 & 0 &  m_{\widetilde{C}_3} & 0 & 0 \\
  0 & 0 & 0 &  m_{\widetilde{C}_4} &0 \\
  0 & 0 & 0 & 0 &  m_{\widetilde{C}_5} \\
\end{array}%
\right).
 \ee
To finish this section, we note that in the model under
consideration there are five charginos; and  they are subject of
the future studies.

\section{Neutralino dark  matter}
\label{neudarkmatt}

 In the model under consideration, there are eleven neutralinos $\widetilde{N}_n $ $(
  n=1,..., 11)$,  each of them  is a linear
  combination of eleven $R=-1$ Majorana fermions, i.e.
  \bea
\widetilde{ N}_n &=& N_{1 n} \widetilde{\mathcal{B}} + N_{2 n}
  \widetilde{\mathcal{W}_3} + N_{ 3 n}
  \widetilde{\mathcal{W}_8} + N_{ 4n}\widetilde{\mathcal{X}^*} + N_{5
  n}\widetilde{\mathcal{X}}\crn && +
 N_{6 n} \widetilde{\rho^o_1} +  N_{7n}
  \widetilde{\rho^{o\prime}_1}+
N_{8n}\widetilde{\chi^o_1 }
  + N_{9 n}\widetilde{\chi^{o\prime}_1 } + N_{10 n}\widetilde{\chi^o_2 }
   + N_{11 n}\widetilde{\chi^{o\prime}_2 }
  \eea
  where   $\widetilde{N}_{n}$  are the
  normalized eigenvectors of the neutralino mass matrix (\ref{an4}).
  The question to be addressed is that our
  consideration below comes  with the conditions (
  \ref{an55}), (\ref{an5}) and  $\mathcal{M}_\mathcal{B}<\mathcal{M}_3
  <\mathcal{M}_8<\mathcal{M}_{45}\ll\mu_\chi,\mu_\rho$.
  Assuming that the neutralino  LSP is a Bino-like $\widetilde{N}_1$, we
  should show its  predicted relic density is consistent with the observational
  data. To answer the question, we must calculate cross section for
  neutralino annihilation and compare it with the observational data
  on dark matter by the WMAP experiment ~\cite{wmapd}
    \be  \Omega_{DM} h^2 = (0.1277^{+ 0.0080}_{-0.0079}) - (0.02229 \pm
    0.00073).
    \label{wmap1} \ee
    In (\ref{wmap}), the normalized Hubble expansion rate
    $h=0.73^{+0.04}_{-0.03}$. We adopt the allowed region as
    \be
    0.0895 <  \Omega_{DM} h^2 < 0.1214.
\label{wmap} \ee

    Before calculating, we should note that a
    precise determinations of the relic density requires the
    solution of the Boltzmann equation governing the evolution of
    the number density $ n \equiv n_{\widetilde{N}}$
    \be
    \frac{dn}{dt}= -3\frac{\dot{a}}{a}n- \langle v
    \sigma \rangle \left( n^2-n_{eq}^2\right)
    \ee
    with $\sigma$ is the cross section of the
    $\widetilde{N}_i$'s annihilation and
    $v$ is the relative velocity. The thermal average $\langle v\sigma \rangle
    $ is defined in the usual manner as any other thermodynamic
    quantity. In the early Universe, the species $\widetilde{N}_i$ were
    initially in thermal equilibrium,
    $n_{\widetilde{N}}=n_{\widetilde{N}^{eq}}$.
    When their typical interaction rate $ \Gamma_{\widetilde{N}}$ became
    less than
    Hubble parameter, $\Gamma_{\widetilde{N}} < H$, the annihilation process
    froze out. Sine then their number in comoving volume has
    remained basically constant

 For the present purpose,  we will use
    approximate solution for $x_f \equiv \frac{T_f}{m_{\widetilde{N}}}$
    \be
    x_f^{-1}=\ln\left[\frac{m_\chi}{2\pi^3} \sqrt{\frac{45}{2g_{*}
    G_N}}\langle v\sigma \rangle  \left(
    x_f\right)x_f^{\frac{1}{2}}\right]
    \ee
     where $g_{*}$ stands for the effective energy degrees
    of freedom at the freeze-out temperature $\left( \sqrt{g_{*}}\simeq9\right)$
    and $G_N$ is the Newton constant. Typically one finds that the
    freeze-out point $x_f$ is
     basically very
    small $(\approx \frac{1}{20})$. The relic mass density $\rho_\chi$ at the
    present is given in ~\cite{anh3}
    \be
    \rho_\chi
    = 4.0\times 10^{-40}\left(
    \frac{T_{\widetilde{N}}}{T_{\gamma}}\right)^3 \left( \frac{T_{\gamma}}{2.8^o
    K}\right)^3 g_*^{\frac{1}{2}}\left(\frac{\textrm{GeV}^{-2}}{ax_f+
    \frac{1}{2}b x_f^2}\right)
    \label{an6}\left(\frac {g}{cm^3}\right)\ee
    with the suppression factor
     $\left(\frac{T_{\widetilde{N}}}{T_\gamma}\right) ^3$ $\approx\frac{1}{20}$
     following  from the entropy conservation in a comoving volume. The
     coefficients $a$ and $b$ are determined by
     \bea
     a&=&\sum_f \theta\left(m_{\widetilde{N}}-m_f\right)\frac{1}{2\pi}
     \frac{p}{m_{\widetilde{N}}}m_f^2\left(A_f -B_f\right)^2, \crn
     b&=&\sum_f \theta\left(m_{\widetilde{N}}-m_f\right)\frac{1}{2\pi}
     \frac{p}{m_{\widetilde{N}}}
     \left[ \left(A_f^2+B_f^2 \right)\left(4m_{\widetilde{N}}^2
     -m_f^2 \right)+6A_f B_f m_f^2 \right]\label{an7}
     \eea
where $p = \sqrt{\left( M_{\widetilde{N}}^2-m_f^2\right)}$ and
$A_f$ and $B_f$ will be defined below.  The sum is taken over the
different types of particle-antiparticle pairs into which the
$\widetilde{N}$ annihilate.

      In order to calculate the LSP mass density,  to
     determine the  $A_f$ and $B_f$ coefficients, we need to
     write down the low-energy effective Lagrangian from interactions.
     The calculation of the
     annihilation cross section in our
      model is straightforward in principle but quite complicate in
      practice.  To ease our work,  we consider only the most important channels
      for  neutralino annihilation in the lowest order
      (tree-level) of perturbation theory for the case
      in which the LSP is a nearly pure Bino $\widetilde{ N}_1$.
      The most important channels
      are annihilation into a pair of fermions
      \be
     \widetilde{ N}_1\widetilde{ N}_1\rightarrow f \widetilde{f},
     (f= q,l,\nu)
      \ee
and into a pair of charged Higgs scalar \be
     \widetilde{ N}_1\widetilde{ N}_1 \rightarrow  H^{+} H^{-}, H^0H^0.\ee

     Because the Bino does not couple to $W^{\pm}$, $Z$  and $ Z^\prime$,
     there is no annihilation of pure Bino to $W^+
 W^-$ and  $Z Z, Z^\prime Z$ or to $Z^\prime Z^\prime $.

    Now we list the couplings  needed in  computation of the
    annihilation cross sections. The couplings of Bino
    $\widetilde{B}$ to quarks and leptons and their two
    scalar partners are given by the following piece of
    Lagrangian:
    \bea
    &-&
\frac{ig^{\prime}}{\sqrt{3}}\left[
-\frac{1}{3}\left(\bar{L}\tilde{L}\bar{\widetilde{B}}-
\bar{\tilde{L}}L\widetilde{B}\right)+
\left(\bar{l}^{c}\tilde{l}^{c}\bar{\widetilde{B}}-
\bar{\tilde{l}}^{c}l^{c}\widetilde{B}\right) \right]
   \crn &-& \frac{ig'}{\sqrt{3}} \left[ \left( \frac{1}{3}
\bar{Q}_1\tilde{Q}_1- \frac{2}{3} \bar{u}^{c}_i\tilde{u}^{c}_i+
\frac{1}{3} \bar{d}^{c}_i\tilde{d}^{c}_i- \frac{2}{3}
\bar{u}^{\prime c}\tilde{u}^{\prime c}+ \frac{1}{3} \bar{d}^{\prime
c}_{\beta}\tilde{d}^{\prime c}_{\beta} \right) \bar{\widetilde{B}}
\right. \crn &-& \left. \left( \frac{1}{3}\bar{\tilde{Q}}_1Q_1-
\frac{2}{3} \bar{\tilde{u}}^{c}_iu^{c}_i+ \frac{1}{3}
\bar{\tilde{d}}^{c}_id^{c}_i- \frac{2}{3} \bar{\tilde{u}}^{\prime
c}u^{\prime c}+ \frac{1}{3} \bar{\tilde{d}}^{\prime
c}_{\beta}d^{\prime c}_{\beta} \right) \widetilde{B} \right]. \nn
    \eea

    The couplings of  neutral Higgs and charged Higgs are determined
    in the following terms
\bea &-& \frac{ig^{ \prime}}{ \sqrt{3}} \left[ - \frac{1}{3}
\left(\bar{\tilde{\chi}} \chi \bar{\widetilde{B}}-
\bar{\chi}\tilde{\chi}\widetilde{B}\right) + \frac{1}{3}
\left(\bar{\tilde{\chi}}^{\prime} \chi^{\prime}
\bar{\widetilde{B}}- \bar{\chi}^{\prime}\tilde{\chi}^{
\prime}\widetilde{B}\right) \right. \crn &&\left.+\frac{2}{3}
\left(\bar{\tilde{\rho}} \rho \bar{\widetilde{B}}-
\bar{\rho}\tilde{\rho}\widetilde{B}\right) - \frac{2}{3}
\left(\bar{\tilde{\rho}}^{\prime} \rho^{\prime}
\bar{\widetilde{B}}-
\bar{\rho}^{\prime}\tilde{\rho}^{\prime}\widetilde{B}\right)
\right]. \eea

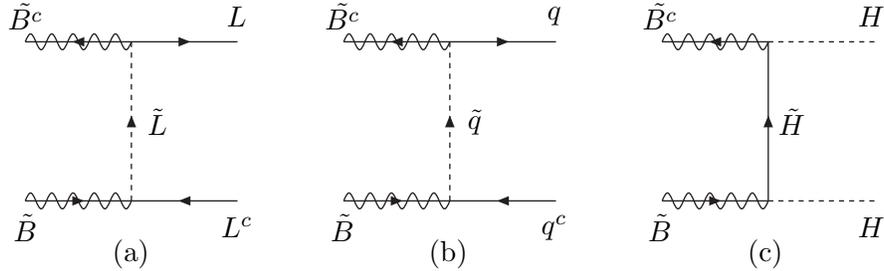
\begin{figure}[htbp]
\bc
\begin{picture}(300,150)(0,-20)

\ArrowLine(40,70)(0,70) \Photon(40,70)(0,70){3}{5}
\ArrowLine(40,70)(80,70) \Text(80,80)[]{$L$}

\DashArrowLine(40,10)(40,70){2} \Text(0,80)[]{$\tilde{B^c}$}

\ArrowLine(0,10)(40,10)

\ArrowLine(80,10)(40,10) \Text(80,0)[]{$L^c$}
 \Photon(0,10)(40,10){3}{5}

 \Text(0,0)[]{$\tilde{B}$}
\Text(50,40)[]{$\tilde{L}$} \Text(40,-10)[]{(a)}

\ArrowLine(160,70)(120,70) \Photon(160,70)(120,70){3}{5}
\ArrowLine(160,70)(200,70) \Text(200,80)[]{$q$}

\DashArrowLine(160,10)(160,70){2} \Text(120,80)[]{$\tilde{B^c}$}

\ArrowLine(120,10)(160,10)

\ArrowLine(200,10)(160,10) \Text(200,0)[]{$q^c$}
 \Photon(120,10)(160,10){3}{5}

 \Text(120,0)[]{$\tilde{B}$}
\Text(170,40)[]{$\tilde{q}$} \Text(160,-10)[]{(b)}

\ArrowLine(280,70)(240,70) \Photon(280,70)(240,70){3}{5}

\DashLine(280,70)(320,70){2} \Text(320,80)[]{$H$}

\ArrowLine(280,10)(280,70) \Text(240,80)[]{$\tilde{B^c}$}

\ArrowLine(240,10)(280,10)

\DashLine(320,10)(280,10){2} \Text(320,0)[]{$H$}
 \Photon(240,10)(280,10){3}{5}

 \Text(240,0)[]{$\tilde{B}$}
\Text(290,40)[]{$\tilde{H}$} \Text(280,-10)[]{(c)}
\end{picture}
\ec
 \caption[]{Feynman diagrams contributing
  to annihilation of Bino  dark matter   }
\label{fig1}
\end{figure}

With the help of the mentioned couplings, the Feynman diagrams for
Bino annihilation processes are depicted in Fig. \ref{fig1}

\vs

We note that the LSP can
      annihilate to  the particles only if theirs mass is lighter than the LSP mass.
      In the~\cite{higph}, by studying the Higgs sector,  we have obtained one charged
      Higgs with mass equal to the W-gauge bosons  mass $\left( m_W \right)$
      and the other ones have
      mass equal to the bilepton mass $M_Y > 440 $ GeV.
      Therefore, in the region $m_{\widetilde{N}} <
      m_W$, the
      LSP cannot annihilate to charged Higgs and  the top-quark as well as the exotic
      quarks and only the annihilation channels into ordinary
      fermion pairs such as  $\widetilde{N} \widetilde{N} \rightarrow f\overline{f}$,
      except
      for the top-quark, are  available.

      From the Feynman  diagram for Bino annihilation processes, the effective
      Lagrangian for a Majorana fermion $\widetilde{N}$
       interacts  with an ordinary quark or lepton $f$ can be
      written down:
      \be
      L_{eff}=\sum_f \overline{\widetilde{N} }  \gamma^\mu \gamma_5
      \widetilde{N}\overline{ f } \gamma_\mu
    \left(
      A_f P_L+B_fP_R\right)f \ee
       with \bea
            A_f&=&\frac{Y_{f_L}^2g^{\prime 2}}{12m_{\widetilde{f}_L}^2}
            -\frac{Y_{f_R}^2g^{\prime
            2}}{12m_{\widetilde{f}_R}^2}, \crn
            B_f &=& -\frac{Y_{f_L}^2g^{\prime 2}}{12m_{\widetilde{f}_L}^2}
            -\frac{Y_{f_R}^2g^{\prime
            2}}{12m_{\widetilde{f}_R}^2}\label{an8}
             \eea
       where $Y_L, Y_R$ are hypercharge of left- and right-handed ordinary quark and
       lepton.

\begin{figure*}[htbp]
\includegraphics[width=12cm,height=12cm]{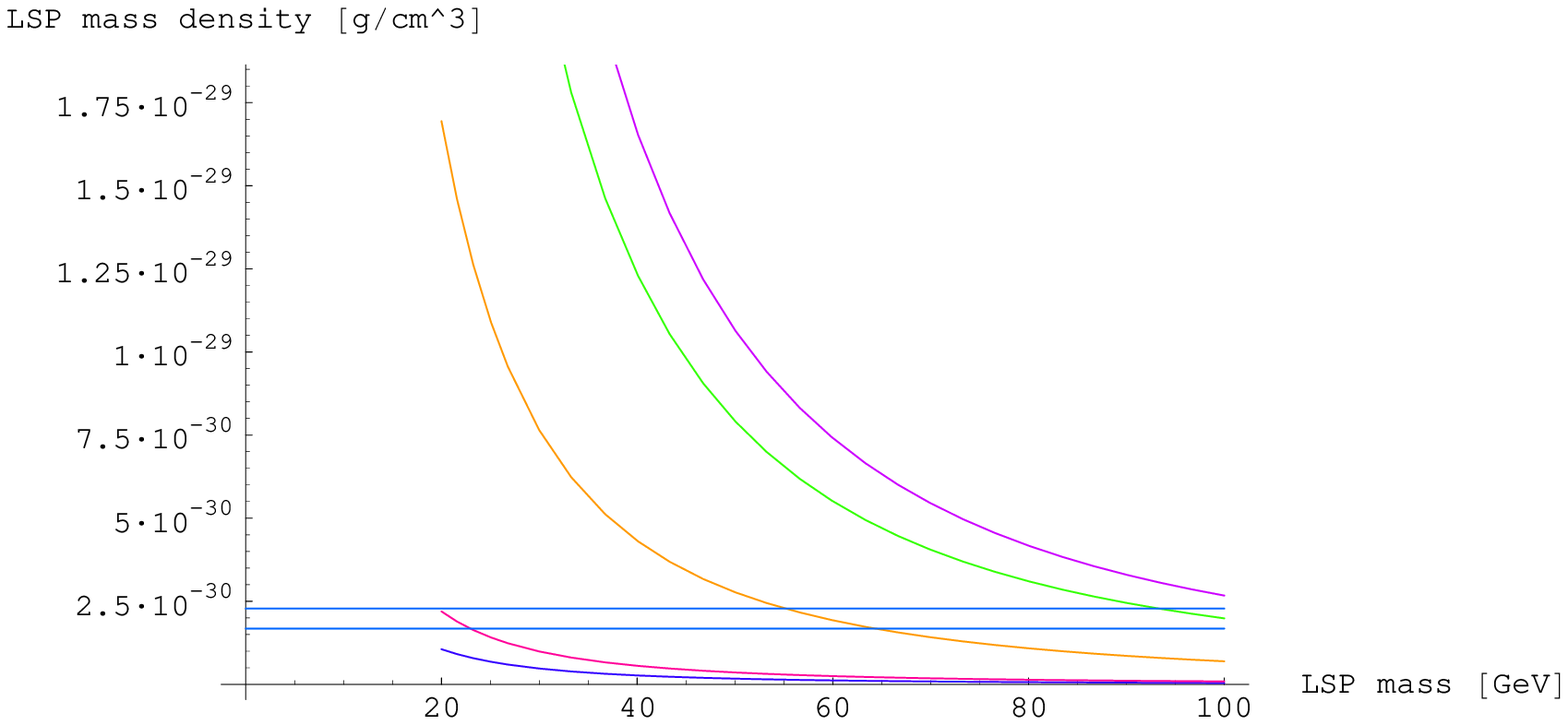}
\caption{\label{lspdensity1}} LSP's mass density as a function of
its mass. The blue, red, yellow, green, violet  curves are allowed
by \
     $m_{\widetilde{f}}=50, 60, 100, 160$ GeV, respectively .
     The horizontal lines are upper and lower
experimental limits given in \cite{wmapd}.
\end{figure*}

    In dealing with Eq.(\ref{an6}),  we have taken into account
      $g^\prime=0.6$ in the model under consideration
     and suggested  that all squarks mass are
     heavier than all sleptons and especially,
     $m_{\widetilde{q}}= 5 m_{\widetilde{l}}$.
     In  Fig. \ref{lspdensity1},  the LSP mass
    density dependence  on its mass has been plotted
        \vs

 From Eqs. (\ref{an6}), (\ref{an7}) and (\ref{an8}),
  it follows that the density increases
 for increasing of  sfermion mass $(\propto m^2_{\tilde{f}})$
 and decreasing of the  LSP mass
 $(\propto \fr{1}{m_{\tilde{N}}})$.
  Fig. \ref{lspdensity1} shows also that the LSP mass is in the
range of 100 GeV.

\begin{figure*}[htbp]
\includegraphics[width=12cm,height=12cm]{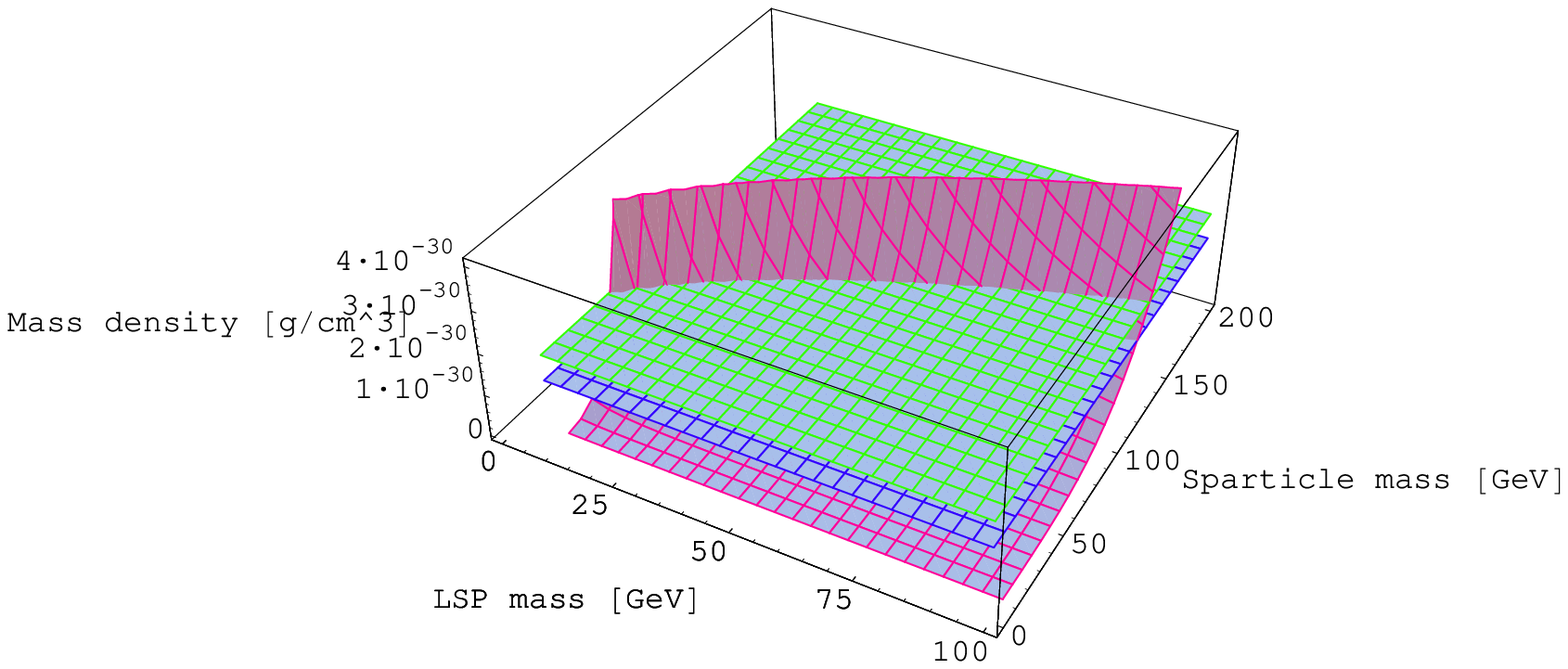}
\caption{\label{lspdensity}} LSP's mass density as a function of
its mass and sparticle's one (grid red plane). The grid green
plane and grid blue plane correspond to the bounds given in
(\ref{wmap1}).
\end{figure*}

In Fig. \ref{lspdensity},
 the LSP mass
    density dependence  on two
    dimensional space of parameters $LSP$ mass and sparticle mass
has been plotted.
      The LSP density
    is drawn as plane.
  We have divided  the space of parameters into allowed and
    disallowed regions, where boundaries of acceptable region
    according to (\ref{wmap})
    are drawn as grid green plane and grid blue plane.
    From the Fig. \ref{lspdensity}, we obtain the lighter sfermion mass is heavier
    than Bino mass.  We also obtain the bounds for mass
    of the sfermions: $60\
     \textrm{GeV}<  m_{\widetilde{f}} < 130
    \ \textrm{GeV}$, while  the masses of the LSP is in the range of:
    $20\ \textrm{GeV} < m_{\widetilde{N}}  < 100\ \textrm{GeV} $.
    It should be noted that this result coincides with estimation
    given in \cite{pdg} (see Fig 1 in page 1114).

\begin{figure*}[htbp]
\includegraphics[width=12cm,height=12cm]{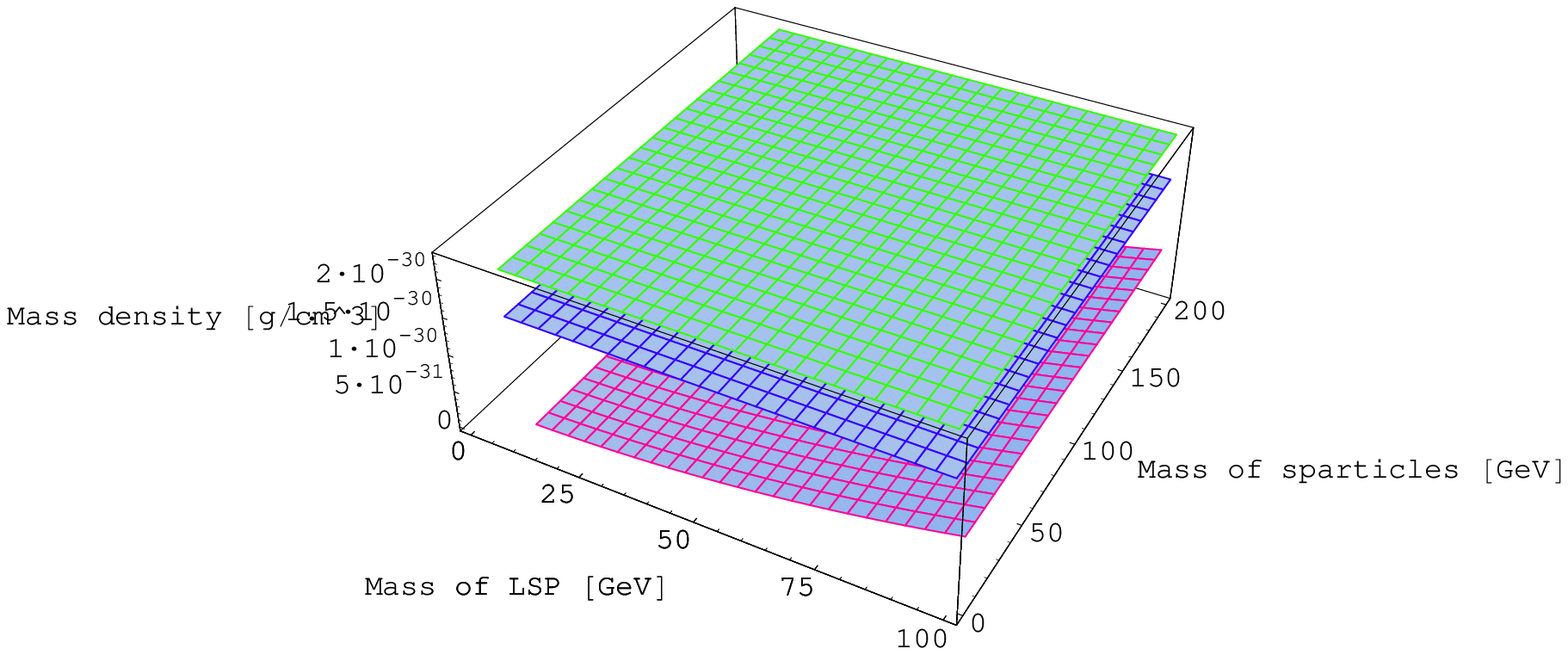}
\caption{\label{lspdensity3}} LSP's mass density as a function of
its mass and sparticle's one (red plane) in the case
$m_{\widetilde{l}}=m_{\widetilde{\widetilde{N}}}$. The grid green
plane and grid blue plane correspond to the bounds given in
\cite{wmapd}.
\end{figure*}

Let us consider the case $ m_{\widetilde{B}}
    =m_{\widetilde{f}}$. The LSP mass density  has been  plotted
in Fig. (\ref{lspdensity3}). The figure shows that
    the LSP mass density is very small;   it is even smaller than the lower
    bound given by the \cite{wmapd}.
     This means that this case is excluded by the WMAP data.

\section{\label{concl} Conclusions}

In this paper we have investigated  the neutralinos and charginos
sector in the supersymmetric economical 3-3-1 model. Accepting
conversational assumption such as in the MSSM,  eigenmasses and
eigenstates in the neutralinos sector were derived. By some
circumstance, the LSP is Bino-like state.

In the charginos sector, the mass matrix can be diagonalized by
two $5 \times 5$ matrices $V$ and $U$.

Assuming that Bino-like is dark matter, its mass density is
calculated.

The cosmological dark matter density gives a bound on  mass  of
LSP neutralino is  in the range of 20 $\div$ 100 GeV. In addition
we have also got a bound on sfermion masses to be: 60 $\div$ 130
GeV. We have also shown that the case $ m_{\widetilde{B}}
    =m_{\widetilde{f}}$ is excluded by the recent experimental WMAP
    data.
 Our result is favored the present bound and it
should be more cleared in the near future. As in the MSSM, the
neutralinos in our model gain the masses in the working region of
the LHC. Consequently they could be checked in coming years.

\section*{Acknowledgments}
The work was supported in part by National Council for Natural
Sciences of Vietnam under grant  No: 402206.
\\[0.3cm]


\begin{thebibliography}{999}

\bibitem{ppf} F. Pisano and V. Pleitez, {\it An $\mathrm{SU}(3)\otimes \mathrm{U}(1)$
model for electroweak interactions}, {\it Phys. Rev.}  {\bf D 46}
(1992) 410; P.H. Frampton, {\it Chiral dilepton model and the
flavor question}, {\it Phys. Rev. Lett.} {\bf 69} (1992) 2889; R.
Foot et al., {\it Lepton masses in an $\mathrm{SU}(3)_L\otimes
\mathrm{U}(1)_N$ gauge model}, {\it Phys. Rev.} {\bf D 47} (1993)
4158.

\bibitem{flt} M. Singer, J.W.F. Valle and J. Schechter, {\it
Canonical neutral-current predictions from the
weak-electromagnetic gauge group SU(3)$\otimes$U(1)}, {\it Phys.
Rev.} {\bf D 22} (1980) 738.

\bibitem{331rh} R. Foot, H.N. Long and Tuan A. Tran,
{\it $\mathrm{SU}(3)_L\otimes\mathrm{U}(1)_N$ and
$\mathrm{SU}(4)_L\otimes\mathrm{U}(1)_N$ gauge models with
right-handed neutrinos}, {\it Phys. Rev.} {\bf D 50} (1994) 34;
J.C. Montero et al., {\it Neutral currents and
Glashow-Iliopoulos-Maiani mechanism in SU(3)$_L\otimes$U(1)$_N$
models for electroweak interactions}, {\it Phys. Rev.} {\bf D 47}
(1993) 2918; H.N. Long, {\it $\mathrm{SU}(3)_L\otimes
\mathrm{U}(1)_N$ model for right-handed neutrino neutral
currents}, {\it Phys. Rev.} {\bf D 54} (1996) 4691; H.N. Long,
{\it $\mathrm{SU}(3)_C\otimes \mathrm{SU}(3)_L\otimes
\mathrm{U}(1)_N$ model with right-handed neutrinos}, {\it Phys.
Rev.} {\bf D 53} (1996) 437.

\bibitem{neu331} Y. Okamoto and M. Yasue, {\it Radiatively generated neutrino
masses in $\mathrm{SU}(3)_L\otimes\mathrm{U}(1)_N$ gauge models},
{\it Phys. Lett.} {\bf B 466} (1999) 267; T. Kitabayshi and M.
Yasue, {\it Radiatively induced neutrino masses and oscillations
in an $\mathrm{SU}(3)_L\otimes\mathrm{U}(1)_N$ gauge model}, {\it
Phys. Rev.} {\bf D 63} (2001) 095002; {\it Two-loop radiative
neutrino mechanism in an $\mathrm{SU}(3)_L\otimes\mathrm{U}(1)_N$
gauge model}, {\it Phys. Rev.} {\bf D 63} (2001) 095006; {\it The
interplay between neutrinos and charged leptons in the minimal
$\mathrm{SU}(3)_L\otimes\mathrm{U}(1)_N$ gauge model}, {\it Nucl.
Phys.} {\bf B 609} (2001) 61; {\it $S_{2L}$ permutation symmetry
for left-handed $\mu$ and $\tau$ families and neutrino
oscillations in an $\mathrm{SU}(3)_L\otimes\mathrm{U}(1)_N$ gauge
model}, {\it Phys. Rev.} {\bf D 67} (2003) 015006; J.C. Montero,
C.A.de S. Pires and V. Pleitez, {\it Neutrino masses through the
seesaw mechanism in 3-3-1 models}, {\it Phys. Rev.} {\bf D 65}
(2002) 095001; A.A. Gusso, C.A.de S. Pires and P.S. Rodrigues da
Silva, {\it Neutrino Mixing and the Minimal 3-3-1 Model}, {\it
Mod. Phys. Lett.} {\bf A 18} (2003) 1849; I. Aizawa et al., {\it
Bilarge neutrino mixing and $\mu$-tau permutation symmetry for
two-loop radiative mechanism}, {\it Phys. Rev.} {\bf D 70} (2004)
015011; A.G. Dias, C.A.de S. Pires and P.S. Rodriguez da Silva,
{\it Naturally light right-handed neutrinos in a 3–3–1 model},
{\it Phys. Lett.} {\bf B 628} (2005) 85; D. Chang and H.N. Long,
{\it Interesting radiative patterns of neutrino mass in an
$\mathrm{SU}(3)_C\otimes\mathrm{SU}(3)_L\otimes\mathrm{U}(1)_X$
model with right-handed neutrinos}, {\it Phys. Rev.} {\bf D 73}
(2006) 053006; P.V. Dong, H.N. Long and D.V. Soa, {\it Neutrino
masses in the economical 3-3-1 model}, {\it Phys. Rev.} {\bf D 75}
(2007) 073006; F. Yin, {\it Neutrino mixing matrix in the 3-3-1
model with heavy leptons and $A_4$ symmetry}, {\it Phys. Rev.}
{\bf D 75} (2007) 073010.

\bibitem{chargequan} C.A.de S. Pires and O.P. Ravinez, {\it Electric
charge quantization in a chiral bilepton gauge model}, {\it Phys.
Rev.} {\bf D 58} (1998) 035008; A. Doff and F. Pisano, {\it Charge
quantization in the largest leptoquark-bilepton chiral electroweak
scheme}, {\it Mod. Phys. Lett.} {\bf A 14} (1999) 1133; {\it
Quantization of electric charge, the neutrino, and generation
universality}, {\it Phys. Rev.} {\bf D 63} (2001) 097903; P.V.
Dong and H.N. Long, {\it Electric Charge Quantization in
$\mathrm{SU}(3)_C\otimes\mathrm{SU}(3)_L\otimes\mathrm{U}(1)_X$
Models}, {\it Int. J. Mod. Phys.} {\bf A 21} (2006) 6677.

\bibitem{CP331} J.T. Liu and D. Ng, {\it Lepton-flavor-changing
processes and CP violation in the
$\mathrm{SU}(3)_c\otimes\mathrm{SU}(3)_L\otimes\mathrm{U}(1)_X$
model}, {\it Phys. Rev.} {\bf D 50} (1994) 548; J.T. Liu, {\it
Generation nonuniversality and flavor-changing neutral currents in
the
$\mathrm{SU}(3)_c\otimes\mathrm{SU}(3)_L\otimes\mathrm{U}(1)_X$
model}, {\it Phys. Rev.} {\bf D 50} (1994) 542; H.N. Long, L.P.
Trung and V.T. Van, {\it Rare Kaon Decay $K^+ \rightarrow \pi^+
\nu \bar{\nu}$ in $\mathrm{SU}(3)_C \otimes \mathrm{SU}(3)_L
\otimes \mathrm{U}(1)_N$ Models}, {\it J. Exp. Theor. Phys.} {\bf
92} (2001) 548, {\it Eksp. Teor. Fiz.} {\bf 119} (2001) 633; J.A.
Rodriguez and M. Sher, {\it Flavor-changing neutral currents and
rare B decays in 3-3-1 models}, {\it Phys. Rev.} {\bf D 70} (2004)
117702; C. Promberger, S.S. Schatt and F. Schwab, {\it
Flavor-changing neutral current effects and CP violation in the
minimal 3-3-1 model}, {\it Phys. Rev.} {\bf D 75} (2007) 115007.

\bibitem{ponce}  W.A. Ponce, Y. Giraldo and L.A. Sanchez,
{\it Minimal scalar sector of 3-3-1 models without exotic electric
charges}, {\it Phys. Rev.} {\bf D 67} (2003) 075001.

\bibitem{haihiggs} P.V. Dong, H.N. Long, D.T. Nhung and D.V.
Soa, {\it $\mathrm{SU}(3)_C \otimes \mathrm{SU}(3)_L \otimes
\mathrm{U}(1)_X$ model with two Higgs triplets}, {\it Phys. Rev.}
{\bf D 73} (2006) 035004.

\bibitem{ahep}  P. V. Dong and H. N. Long,  {\it The economical
$\mathrm{SU}(3)_C\otimes \mathrm{SU}(3)_L \otimes \mathrm{U}(1)_X$
model}, [arXiv:0804.3239(hep-ph)](2008),
 to appear  in  {\it Advances in High Energy Physics}.

\bibitem{higgseconom} P.V. Dong, H.N. Long and D.V. Soa, {\it Higgs-gauge boson
interactions in the economical 3-3-1 model}, {\it Phys. Rev.} {\bf
D 73} (2006) 075005.

\bibitem{dlhh} P.V. Dong, T.T. Huong, D.T. Huong and H.N. Long,
{\it Fermion masses in the economical 3-3-1 model}, {\it Phys.
Rev.} {\bf D 74} (2006) 053003.

\bibitem{dls1} P.V. Dong et al., in Ref. \cite{neu331}.

\bibitem{nogo} S. Coleman and J. Mandula, {\it
All Possible Symmetries of the S Matrix},
{\it Phys. Rev.} {\bf 159} (1967) 1251.

\bibitem{susy} See, for example, J. Wess and J. Bagger,
{\it Supersymmetry and Supergravity}, 2nd edition, Princeton
University Press, Princeton NJ, (1992); H.E. Haber and G.L. Kane,
{\it The search for supersymmetry: Probing physics beyond the
standard model}, {\it Phys. Rep.} {\bf 117} (1985) 75.

\bibitem{martin} S. Martin, {\it A supersymmetry primer}, [arXiv:hep-ph/9709356].

\bibitem{SIDM} V. Silveira and A. Zee, {\it Scalar Phantoms},
{\it Phys. Lett.} {\bf B 161} (1985) 136; D.E. Holz and A. Zee,
{\it Collisional dark matter and scalar phantoms}, {\it Phys.
Lett.} {\bf B 517} (2000) 239;  C.P. Burgess, M. Pospelov and T.
ter Veldhuis, {\it The Minimal Model of nonbaryonic dark matter: a
singlet scalar}, {\it Nucl. Phys.} {\bf B 619} (2002) 709; B.C.
Bento, O. Bertolami, R. Rosenfeld and L. Teodoro, {\it
Self-interacting dark matter and the Higgs boson}, {\it Phys.
Rev.} {\bf D 62} (2000) 041302;  J. McDonald, {\it Gauge singlet
scalars as cold dark matter}, {\it Phys. Rev.} {\bf D 50} (1994)
3637; {\it Thermally Generated Gauge Singlet Scalars as
Self-Interacting Dark Matter}, {\it Phys. Rev. Lett.} {\bf 88}
(2002) 091304.

\bibitem{ss} D. N. Spergel and P. J. Steinhardt, {\it Observational Evidence for
Self-Interacting Cold Dark Matter}, {\it Phys. Rev. Lett.} {\bf
84} (2000) 3760.

\bibitem{msusy} J.C. Montero, V. Pleitez, M.C. Rodriguez,
{\it Supersymmetric 3-3-1 model}, {\it Phys. Rev.} {\bf D 65}
(2002) 035006.

\bibitem{duongma}  T.V. Duong and E. Ma, {\it Supersymmetric $\mathrm{SU}(3)
\otimes  \mathrm{U}(1)$ gauge models: Higgs structure at the
electroweak energy scale}, {\it Phys. Lett.} {\bf B 316} (1993)
307; {\it  Scalar mass bounds in two supersymmetric extended
electroweak gauge models}, {\it J. Phys} {\bf G 21} (1995) 159;
M.C. Rodriguez, {\it Scalar sector in the minimal supersymmetric
3-3-1 model}, {\it Int. J. Mod.  Phys.} {\bf  A 21} (2006) 4303.

\bibitem{leptonmassm331}J.C. Montero, V. Pleitez and  M.C.
Rodriguez, {\it Lepton masses in a supersymmetric 3-3-1 model},
{\it Phys. Rev.} {\bf D 65} (2002) 095008; C.M. Maekawa and M.C.
Rodriguez, {\it Masses of fermions in supersymmetric models}, {\it
JHEP} {\bf 04} (2006) 031.

\bibitem{consm331}  M. Capdequi-Peyranere, M.C. Rodriguez, {\it
Charginos and neutralinos production at 3-3-1 supersymmetric model
in $e^- e^-$ scattering}, {\it Phys. Rev.} {\bf D 65} (2002)
035001.
\bibitem{longpal}  Hoang Ngoc Long and Palash B Pal, {\it
Nucleon instability in a supersymmetric $SU(3)_C \otimes SU(3)_L
\otimes U(1)$ model}, {\it Mod. Phys. Lett.} {\bf A 13} (1998)
2355.

\bibitem{s331r} J.C. Montero, V. Pleitez and M.C. Rodriguez, {\it Supersymmetric
3-3-1 model with right-handed neutrinos}, {\it Phys. Rev.} {\bf D
70} (2004) 075004.

\bibitem{scalarrhn}  D.T. Huong,  M.C. Rodriguez and  H.N.
Long, {\it Scalar sector of supersymmetric $SU(3)_C \otimes$ $
SU(3)_L \otimes $ $ U(1)_N$  model with right-handed neutrinos},
[arXiv:hep-ph/0508045].

\bibitem{marcos} P.V. Dong, D.T. Huong, M.C. Rodriguez and H.N.
Long, {\it Neutrino masses in the supersymmetric $\mathrm{SU}(3)_C
\otimes \mathrm{SU}(3)_L \otimes \mathrm{U}(1)_X$ model with
right-handed neutrinos}, {\it Eur. Phys. J.} {\bf C 48} (2006)
229.

\bibitem{nonst331} R. Martinez,  William A. Ponce and Luis A.
Sanchez, {\it $SU(3)_c \otimes$ $ SU(3)_L \otimes $ $ U(1)_X$ as
an $E_6$ subgroup}, {\it Phys. Rev.} {\bf D 64} (2001) 075013;
David L. Anderson and  Marc Sher, {\it 3-3-1 models with unique
lepton generations}, {\it Phys. Rev.} {\bf D 72} (2005) 095014.

\bibitem{consnst}  R.  A. Diaz, R. Martinez, J. Alexis Rodriguez ,
{\it A new supersymmetric $\mathrm{SU}(3)_L \otimes
\mathrm{U}(1)_X$ gauge model}, {\it Phys. Lett.} {\bf B 552}
(2003) 287.

\bibitem{susyeco} P.V. Dong, D.T. Huong, M.C. Rodriguez and H.N.
Long, {\it Supersymmetric economical 3-3-1 model}, {\it Nucl.
Phys.} {\bf B 772} (2007) 150.

\bibitem{higph} P. V. Dong, D. T. Huong, N. T. Thuy and H. N. Long,
{\it Higgs phenomenology of supersymmetric economical 3-3-1 model,
Nucl. Phys}.  {\bf B 795} (2008) 361.


\bibitem{jhep}P. V. Dong, Tr. T. Huong, N. T. Thuy and H. N. Long,
{\it Sfermion masses  in the supersymmetric economical 3-3-1
model},  {\it JHEP} {\bf 11} (2007) 073.


\bibitem{changlong} D. Chang and H.N. Long, in Ref. \cite{neu331};
See also, M.B. Tully and G.C. Joshi, {\it Generating neutrino mass
in the 3-3-1 model}, {\it Phys. Rev.} {\bf D 64} (2001) 011301.

\bibitem{wmapd} D. N. Spergel {\it et al.} [WMAP Collaboration],
{\it Wilkinson Microwave Anisotropy Probe (WMAP) three year results:
implications for cosmology.  Astrophys. J. Suppl.} {\bf 170}, 377
(2007).

\bibitem{anh3}J.  Ellis  {\it  et al, Supersymmetric relics from
the Big Bang,  Nucl. Phys}. {\bf B 238} (1984) 453.

\bibitem{pdg} Particle Data Group collaboration, W.-M. Yao et. al.,
{\it Review of particle physics}, {\it J. Phys.} {\bf G 33} (2006)
1, p. 114.
\end{thebibliography}
\end{document}